\documentclass[aps,floats,showpacs]{revtex4}
\usepackage{graphicx}
\usepackage{epsfig}
\newcommand{\bastar}{\begin{eqnarray*}}
\newcommand{\eastar}{\end{eqnarray*}}
\newskip\humongous \humongous=0pt plus 1000pt minus 1000pt

\newif\ifdtup

\relax
\newcommand{\be}{\begin{equation}}
\newcommand{\ee}{\end{equation}}
\newcommand{\bea}{\begin{eqnarray}}
\newcommand{\eea}{\end{eqnarray}}
\newcommand{\pro}{\partial}

\newcommand{\dfrac}{\displaystyle\frac}
\newcommand{\ba}{\begin{array}}
\newcommand{\ea}{\end{array}}
\newcommand{\nn}{\nonumber}

\newcommand{\rt}{{\vec r}}
\newcommand{\cH}{{\cal H}}
\newcommand{\cE}{{\cal E}}

\begin{document}
\title{QCD effective action with
a most general homogeneous field background}
\bigskip
\author{Y.M. Cho}
\email{ymcho@yongmin.snu.ac.kr}
\author{J.H. Kim}
\affiliation{Center for Theoretical Physics
and School of Physics, \\
College of Natural Sciences,
Seoul  National University, Seoul 151-742, Korea}
\author{D.G. Pak}
\email{dmipak@phya.snu.ac.kr}
\affiliation{ Center for Theoretical Physics,
Seoul National University, Seoul 151-742, Korea}
\affiliation{Institute of Applied Physics,
Uzbekistan National University, Tashkent 700-095, Uzbekistan}
\bigskip
\begin{abstract}
We consider one-loop effective action of $SU(3)$ QCD with a most
general constant chromomagnetic (chromoelectric) background which
has two independent Abelian field components.
The effective potential with a pure magnetic
background has a local minimum only when two Abelian components
$H_{\mu\nu}^3$ and $H_{\mu\nu}^8$ of color magnetic field are
orthogonal to each other. The non-trivial structure of the effective
action has important implication in estimating quark-gluon
production rate and $p_T$-distribution in quark-gluon plasma. In
general the production rate depends on three independent Casimir
invariants, in particular, it depends on the relative orientation
between chromoelectric fields.
\end{abstract}
\maketitle

{\bf 1. Introduction}
\vskip 2mm

An interesting problem which has been studied recently is to
calculate the soft gluon production rate in a constant
chromoelectric background \cite{cash,nayak}. This problem arises
when one wishes to find the gluon production rate in quark-gluon
plasma produced in high energy hadron collider experiments
\cite{exp}. One faces the same problem when one wants to estimate
the decay rate of a chromoelectric knot which might exist in QCD
\cite{chopl}. This problem is closely related to the problem to
calculate the QCD effective action in a constant chromoelectric
background since the imaginary part of the effective action
determines the production rate \cite{schw}. There have been
considerable amount of discussions on this problem in the literature
\cite{savv,yil,sch,ambj1,ditt,flyv,avr,prd02,cho99}.

In the present Letter we calculate one-loop QCD
effective action in most general homogeneous
chromomagnetic and chromoelectric external fields.
The generic structure of $SU(3)$ QCD effective action is
not much different from that of $SU(2)$ QCD.
A new feature is that $SU(3)$ Lie algebra has rank
two due to the Cartan subalgebra $U(1)\times U(1)$.
This implies that the most general homogeneous
chromomagnetic (or chromoelectric) field
contains two independent vector fields directed along
two Abelian directions in the internal color space,
or equivalently, along two directions in the configuration space.
This leads to a more non-trivial structure of the
effective action with a most general constant background, and that is
what should be taken into account when solving some physical problems.
Specifically, the real part of the effective potential
for a constant color magnetic background
has a local minimum only when two background chromomagnetic
vector fields are orthogonal to each other. This unexpected
and surprising result had been obtained first by Flyvbjerg \cite{flyv}
who suggested an improvement of the Copenhagen vacuum.
Another implication is that the quark-gluon production rate
in a most general chromoelectric external background depends on
the angle between two independent chromoelectric vector
fields. That means the quark-gluon production rate
depends on three Casimir invariants in general.

{\bf 2. Effective action}

We consider a constant field background which can be defined in an
appropriate gauge by only Abelian gauge components
$A_{\mu}^i~(i=3,8)$ corresponding to the Cartan algebra of $SU(3)$.
To calculate the effective action we integrate out the off-diagonal
(valence) gluons $X_{\mu}^a$ from the generating functional of
one-point irreducible Green functions. For this it is convenient to
introduce three complex vector fields ($W_\mu^p,~p=1,2,3$ )
\bea
W^{1}_{\mu}= \dfrac{1}{\sqrt 2} (X_\mu^1 + i X_\mu^2), ~~~~~
W^{2}_{\mu}= \dfrac{1}{\sqrt 2} (X_\mu^6 + i X_\mu^7), ~~~~~
W^{3}_{\mu}= \dfrac{1}{\sqrt 2} (X_\mu^4 - i X_\mu^5).
\eea
This
allows us to express a pure QCD Lagrangian in an explicitly Weyl
invariant form
\bea
&{\cal L} = -\dfrac{1}{4} \vec F_{\mu\nu}^2
=\sum_p \Big \{-\dfrac{1}{6} ({\cal G}_{\mu\nu}^p)^2 +\dfrac{1}{2}
|D_{p\mu} W_{\nu}^p- D_{p\nu} W_{\mu}^p|^2 -ig {\cal G}_{\mu\nu}^p
W_\mu^{p*} W_\nu^p -\dfrac{1}{2} g^2 \big[(W_\mu^{p*}W_\mu^p)^2
-(W_\mu^{p*})^2 (W_\nu^p)^2 \big]\Big \} , \nn \\
&{\cal G}_{\mu\nu}^p = \partial_\mu {\cal B}_\nu^p-
\partial_\nu {\cal B}_\mu^p, ~~~~~~D_{p\mu}W_\nu^p =
(\partial_\mu - ig {\cal B}_\mu^p) W_\nu^p,  ~~~~~~
{\cal B}_\mu^p=A_\mu^i \rt_{i}^{\,p},
\eea
where the $SU(3)$ root vectors $\rt_p$
are given by
\bea
&&\rt_{i}^{\,1} = (1,0), ~~~~\rt_{i}^{\,2} =
(-1/2,\sqrt 3/2), ~~~~ \rt_{i}^{\,3} = (-1/2,-\sqrt 3/2) .
\eea
Notice that the Abelian background fields ${\cal B}_\mu^p$ are
precisely the dual potentials in $i$-spin, $u$-spin, and $v$-spin
direction in color space which couple to three valence gluons
$W_{\mu}^p$. With this we have the following functional integral
form of the one-loop effective action
\bea
&\exp\Big[iS_{eff}(A_\mu) \Big] = \dfrac{}{} \sum_p \int {\cal D} (W_\mu,
c_1, c_2) \exp \Big{\lbrace} \dfrac{}{} i\int \Big[ -\dfrac{1}{6}
{\cal G}_{\mu\nu}^2
+\dfrac{1}{2}|{D}_\mu{W}_\nu-{D}_\nu{W}_\mu|^2 \nn\\
&-ig {\cal G}_{\mu\nu} W_\mu^* W_\nu
-\dfrac{1}{2} g^2 \Big[(W_\mu^*W_\mu)^2-(W_\mu^*)^2 (W_\nu)^2 \Big]
-\dfrac{1}{\xi} |{D}_\mu W_\mu|^2 \nn\\
&+ c_1^\dagger ({D}^2 + g^2W_\mu^* W_\mu )c_1 - g^2 c_1^\dagger
 W_\mu W_\mu c_2 + c_2^{\dagger}
({D}^2 + g^2W_\mu^* W_\mu )^*c_2 - g^2 c_2^{\dagger} W^*_\mu W^*_\mu
c_1 ~\Big] d^4x \Big{\rbrace},
\eea
where $\xi$ is a gauge fixing
parameter and $c_{1,2}$ are the ghost fields, and
here we have suppressed the summation index $p$ in the integrand.
Now a few remarks are in order. First, notice that except for the
$p$-summation the integral expression is identical to that of
$SU(2)$ QCD \cite{prd02}. This shows that one can reduce the
calculation of QCD effective action to that of $SU(2)$ QCD.
Secondly, the above result can easily be generalized to $SU(N)$ QCD
with $N(N-1)/2$ $p$-summation. Thirdly, one might include the
Abelian part in the functional integration, but this does not affect
the result because the Abelian part has no self-interaction. This
tells that only the valence gluon loops contribute to the
integration.

Now, in the same manner as in $SU(2)$ QCD \cite{prd02} we can derive
the functional determinant form for the one-loop correction $\Delta
S$ to the effective action (with $\xi = 1/2$)
\bea
\Delta S &=& i
\sum \limits_p \ln {\rm Det} [(-D_p^2+2gH_p) (-D_p^2-2gH_p)]
+ i \sum \limits_p \ln {\rm Det} [(-D_p^2-2igE_p)(-D_p^2+2igE_p)] \nn \\
&&- 2i \sum_p \ln {\rm Det}[-D_p^2], \nn \\
H_p &=& \dfrac{1}{2} \sqrt {\sqrt {{\cal G}_p^4
+ ({\cal G}_p \tilde {\cal G}_p)^2} + {\cal G}_p^2},~~~~~~
E_p = \dfrac{1}{2} \sqrt {\sqrt {{\cal G}_p^4
+ ({\cal G}_p \tilde {\cal G}_p)^2} - {\cal G}_p^2}, ~~~~~~
\tilde {\cal G}_{\mu\nu}=\dfrac{1}{2} \epsilon_{\mu\nu\rho\sigma}
               {\cal G}^{\rho\sigma}, \label{eq:det}
\eea
from which with Schwinger's proper time method we obtain
\bea
&\Delta {\cal L} =  \lim \limits_{\epsilon\rightarrow 0}
\dfrac{g^2}{16 \pi^2} \sum_p  \int_{0}^{\infty}
\dfrac{dt}{t^{1-\epsilon}} \dfrac{H_p E_p}{\sinh (gH_p t/\mu^2)
\sin (gE_p t/\mu^2)} \Big[ \exp(-2gH_p t/\mu^2)+\exp(+2gH_p t/\mu^2) \nn\\
&+\exp(+2igE_p t/\mu^2)+\exp(-2igE_p t/\mu^2)-2 \Big],
\label{eq:schw}
\eea
where $\mu$ is a mass parameter.
One should emphasize that the expression
(\ref{eq:det}) is valid for arbitrary magnetic and electric fields,
whereas the integral representation (\ref{eq:schw}) is applicable
only to constant field configurations. Notice also that the integral
representation is intrinsically ill-defined and has ambiguity due to
the pole structure. Moreover, it contains a well-known infra-red
divergence which has to be regularized.
The mathematical ambiguity
in (\ref{eq:schw}) reflects the existence of different physical
problems to which the constant field approximation has been applied.

The case of general electric-magnetic background of $SU(3)$ QCD is
in full analogy with the corresponding case of $SU(2)$ QCD. The
analytical expression for the effective action of $SU(2)$ QCD with a
general electric-magnetic constant background has been obtained in
\cite{cho99}. So that we will consider only two special cases of
pure magnetic and pure electric external field concentrating on
features of the $SU(3)$ structure of QCD.

Let us consider first the constant chromomagnetic external field.
The corresponding effective Lagrangian of $SU(2)$ QCD including both, the real
and imaginary parts, had been calculated in the well-known
paper by Nielsen and Olesen
\cite{savv}. The corresponding expression for $SU(3)$ QCD
has the same structure
\bea
{\cal L}_{eff} &=& - \sum \limits_p \Big(\dfrac{H_p^2}{3}
+\dfrac{11g^2}{48\pi^2} H_p^2(\ln \dfrac{gH_p}{\mu^2}-c)
+ \dfrac{ig^2}{8\pi} H_p^2 \Big), \label{eq:Lmag}
\eea
where $c=1.2921...$ (within the modified minimal subtraction scheme).
The Lagrangian has an imaginary part which
implies the existence of a tachyonic mode in the
theory and instability of the constant external
chromomagnetic field.

The effective Lagrangian possesses a manifest Weyl symmetry provided by
the six-element subgroup of $SU(3)$ which contains
the cyclic group $Z_3$.
We can also express the effective Lagrangian (\ref{eq:Lmag})
in terms of three
Casimir invariants
\bea
C_2=\dfrac{1}{2} (F_{\mu\nu}^i)^2, ~~~~
 C_4=(d^{ijk} F_{\mu\nu}^j F_{\mu\nu}^k)^2, ~~~~
 C_6=(d^{ijk} F_{\mu\nu}^i F_{\nu\rho}^j F_{\rho \sigma}^k)^2.
\eea

One can check that $H_p$ satisfy
the equations:
\bea
&& H_1^2+H_2^2+H_3^2=3 C_2 \equiv \alpha, \nn \\
&& H_1^2 H_2^2 + H_2^2 H_3^2+H_3^2 H_1^2 =
        3 C_2^2-\dfrac{9}{16} C_4 \equiv \beta, \nn  \\
&& H_1^2 H_2^2 H_3^2 = C_2^3-\dfrac{3}{8} C_2 C_4 -\dfrac{3}{2} C_6
 \equiv \gamma, \label{eq:cubeqs}
\eea
where we denote for a convenience the
right hand sides of the equations
by $\alpha,\beta, \gamma$ respectively.

We generalize the equations for $H_p$ (\ref{eq:cubeqs}) by assuming
that arbitrary constant magnetic background fields $H_p $ satisfy
the same equations, so that $H_p^2$ are represented by real roots of
the cubic equation
\bea
&& x^3 - \alpha x^2 +\beta x -\gamma=0.
\eea
The solution to the equation provides the values of $H_p$ in terms
of Casimir invariants
\bea
&& H_p^2 = \dfrac{\sqrt{3 C_4}}{2} \sin \phi_p
+C_2, \label{eq:magsol}
\eea
where $\phi_p$ are three basic
solutions of the equation ($[0\leq \phi \leq 2 \pi]$)
\bea
&& \sin 3
\phi = \dfrac{2}{\sqrt 3} \dfrac{(8 C_6-C_2 C_4)}{C_4^{3/2}}.
\label{cubsol}
\eea

\begin{figure}[t]
\hfill
  \begin{minipage}[t]{.45\textwidth}
\begin{center}
\psfig{figure=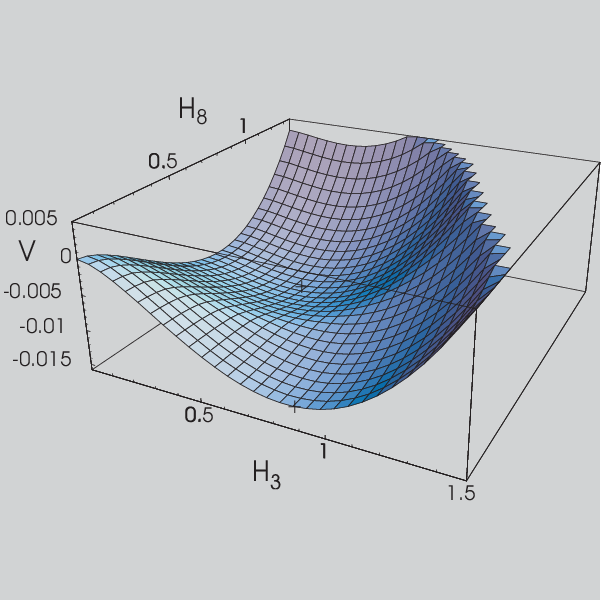, height = 5.5 cm, width = 6.5 cm}
\caption{\label{Fig. 1} The QCD effective potential with $\cos \theta =1$,
which has two degenerate minima ($c=5/4$).}
\end{center}
\end{minipage}
  \hfill
  \begin{minipage}[t]{.45\textwidth}
\begin{center}
\psfig{figure=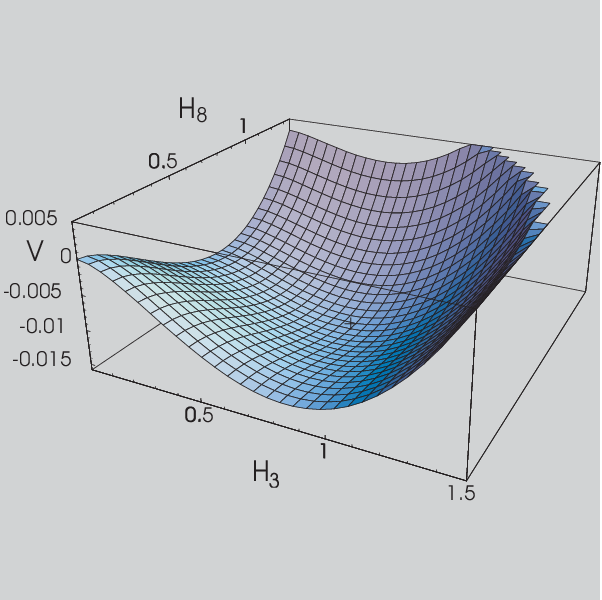, height = 5.5 cm, width = 6.5 cm}
\caption{\label{Fig. 2} The effective potential with $\cos \theta =0$,
which has a unique minimum at $\bar H_3=\bar H_8=H_0$ ~($c=5/4$).}
\end{center}
 \end{minipage}
  \hfill
\end{figure}

Just as in $SU(2)$ QCD we can obtain the effective potential
from the effective action. For the constant magnetic background
a real part of the effective potential is given by
\bea
V_{eff}&=&\dfrac{1}{2} (\bar H_3^2+\bar H_8^2) +\dfrac{11g^2}{48 \pi^2}
\Big \{\bar H_3^2 \big(\ln \dfrac{g\bar H_3}{\mu^2} -c\big)
+\bar H_+^2 \big(\ln \dfrac{g\bar H_+}{\mu^2} -c\big)
+\bar H_-^2 \big(\ln \dfrac{g\bar H_-}{\mu^2} -c\big)\Big \}, \nn \\
\bar H_\pm^2&=&\dfrac{1}{4} \bar H_3^2+\dfrac{3}{4} \bar H_8^2
\pm \dfrac{\sqrt 3}{2} \bar H_3 \bar H_8 \cos \theta, \nn\\
\bar H_3&=& \sqrt{(H_{\mu\nu}^3)^2/2},
~~~\bar H_8= \sqrt{(H_{\mu\nu}^8)^2/2},~~~~
\cos \theta = H_{\mu\nu}^3 H_{\mu\nu}^8/2 \bar H_3 \bar H_8.\label{effpot}
\eea
Notice that the classical potential depends only
on $\bar H_3^2+\bar H_8^2$, but the effective potential
depends on three variables $\bar H_3$, $\bar H_8$,
and $\cos \theta$. We emphasize that $\cos \theta$ can be arbitrary
because $H_{\mu\nu}^3$ and $H_{\mu\nu}^8$ are completely
independent, so that they can have different
space polarization.

When $H_{\mu\nu}^3$ and $H_{\mu\nu}^8$ are parallel ($\cos
\theta=1$) it has two degenerate minima at $\bar H_3=2^{1/3}
H_0,~\bar H_8=0$ and at $\bar H_3=2^{-2/3} H_0,~\bar
H_8=2^{-2/3}\sqrt3 H_0$. When $\theta$ approaches the value
$\pm\pi/2$ the two minima merge into one minimum at $\bar H_3=\bar
H_8=H_0$
\bea
H_0 = \dfrac{\mu^2}{g} \exp \big(-\dfrac{16\pi^2}
{11g^2} +c-\dfrac{1}{2} \big).
\eea
We plot the effective potential
for $\cos \theta=1$ in Fig. 1 and for $\cos \theta =0$ in Fig. 2 for
comparison.

Usually, in most physical applications, the constant background
field is chosen to be directed along one direction in the
configuration (or internal) space by imposing the constraint
$\cos\theta=1$. Our analysis shows that if we start with such a
special background we would never reach the absolute minimum of the
effective potential. This non-trivial feature of the energy
functional for a pure QCD had been found first in \cite{flyv}.
Notice also that this minimum represents a saddle point in the space
of all possible non-constant chromomagnetic fields due to the
presence of the Nielsen-Olesen imaginary part. So that it does not
correspond to a true stable vacuum. A possible stable vacuum,
so-called "Copenhagen vacuum", has been proposed in
\cite{niel2,ambj3}. An interesting example of a stable solution made
of a pair of monopole-antimonopole strings in $SU(2)$ QCD has been
obtained recently in \cite{cho11}.

One can renormalize the potential
by defining a running coupling $\bar g^2(\bar \mu^2)$
\bea
&\dfrac{\pro^2 V_{eff}}{\pro \bar H_i^2}
\Big |_{\bar H_3=\bar H_8=\bar \mu^2,\theta=\pi/2}
        =\dfrac{g^2}{\bar g^2}, \label{renorm}
\eea
from which we can retrieve the correct QCD $\beta$-function. The
renormalized potential has the same form as in (\ref{effpot}), with
the formal replacement $g\rightarrow \bar g,~\mu \rightarrow \bar
\mu,~c=5/4$. It has the unique absolute minimum
\bea
& V_{min} = -\dfrac{11
\bar \mu^4}{32 \pi^2} \exp \big(-\dfrac{32\pi^2} {11 \bar g^2}
+\dfrac{3}{2} \big).
\eea

For a constant chromoelectric field background one can obtain a
similar integral expression for the one-loop contribution to the
effective Lagrangian
\bea
\Delta {\cal L}=\dfrac{g}{16 \pi^2} \lim
\limits_{\epsilon\rightarrow0} \sum_p  \int_{0}^{\infty}
\dfrac{dt}{t^{2-\epsilon}} \dfrac{E_p}{\sin (gE_p t)}
 \Big(\exp(+2igE_p t)+\exp(-2igE_p t) \Big) . \label{eq:intelec}
\eea
The chromoelectric fields $E_p$ can be expressed in terms of
corresponding Casimir invariants by an equation similar to
(\ref{eq:magsol}). In a special case, when two background
chromoelectric fields $F_{\mu\nu i}$ lie along one direction in the
color space, the Casimir invariant $C_4$ is not longer independent
and can be expressed in terms of lower Casimir $C_2$. With this the
general solution for $E_p$ is simplified to a special one obtained
recently in \cite{nayak}.
\vskip 2mm

{\bf 3. Quark production rate}

\vskip 2mm
The quark contribution to the effective action of QCD
does not have strong infra-red divergency problem, and its
calculation is straightforward as in $SU(2)$ theory. The Lagrangian
of $SU(3)$ QCD with quarks interacting with the Abelianized gauge
potential can be written as follows
\bea
&{\cal L}_{q}= -\dfrac{1}{8}
\sum_p ({\cal F}_{\mu\nu}^p)^2 + \sum_p \bar \Psi_p(i\gamma^\mu
{\cal D}_{p\mu} -m) \Psi_p, \nn \\
& {\cal F}_{\mu\nu}^p=\pro_\mu {\cal
A}_\nu^p -\pro_\nu {\cal A}_\mu^p, ~~~~ {\cal D}_{p\mu} = \pro_\mu
-i\dfrac{g}{2} {\cal A}_\mu^p, ~~~~ {\cal A}_\mu^p= A_\mu^i \vec
w_i^{\,p},\label{qlag}
\eea
where $m$ is the quark mass, and $\vec w^{\,p}$
are weights of $SU(3)$. One can express the quark
contribution to the one-loop effective action in a Weyl invariant
form
\bea
\Delta {\cal L}_{q}=-\dfrac{g^2}{16 \pi^2} \sum_p
\int_{0}^{\infty} \dfrac{dt}{t^{1-\epsilon}} \cH_p \cE_p
 \coth (g\cH_p t/2)
 \cot (g\cE_p t/2)\exp(-m^2t) \label{quarklagr},
\eea
where we introduce gauge invariant
variables $\cH_p,~\cE_p$ corresponding to
the pure magnetic and electric fields defined as in (\ref{eq:det}).
The analytical series representation for $SU(2)$ QCD effective action
with a general constant background has been obtained in \cite{cho99}, so that
we will concentrate mainly on some new features appeared
in $SU(3)$ theory and consider particularly the
pair production in quark-gluon plasma in what follows.

Let us consider $p_T$-distribution of the quark production
rate in constant chromoelectric background
\bea
&\Delta {\cal L}_{q}= -\dfrac{g}{16 \pi^3}\sum_p
\int_{0}^{\infty} d^2 p_T \dfrac{dt}{t^{1-\epsilon}}{\cal E}_p
\cot (g {\cal E}_p t/2)
\exp[-(p_T^2+m^2)t], \nn \\
&{\cal E}_1 = {\cal E}_+,~~~{\cal E}_2={\cal E}_-,
~~~{\cal E}_3= 2 \bar E_8/\sqrt 3, \nn\\
&{\cal E}_\pm =\sqrt{\bar E_3^2+\bar E_8^2/3 \pm 2 \bar E_3 \bar E_8
\cos \theta/\sqrt 3}, \nn\\
&\bar E_3 =\sqrt {(F_{\mu\nu}^3)^2/2}, ~~~\bar E_8=\sqrt
{(F_{\mu\nu}^8)^2/2}, ~~~ \cos \theta=F_{\mu\nu}^3
F_{\mu\nu}^8/2\bar E_3 \bar E_8,\label{eq:quark2}
\eea
here, $\theta$ is the angle between two chromoelectric fields
$F_{\mu\nu}^3$ and $F_{\mu\nu}^8$. For the quark contribution we
have no acausal states, so that the contour above the $t$-axis from
$0+\epsilon$ does become the causal contour. This implies
\cite{nayak}
\bea
Im~\Delta {\cal L}_{q}=\dfrac{g}{16 \pi^3} \sum_p
\sum_{n=1}^\infty \dfrac{1}{n} {\cal E}_p \exp \Big (-\dfrac{2\pi n
(p_T^2+m^2)}{g{\cal E}_p}\Big). \label{eq:quark3}
\eea
The imaginary
part depends on three independent variables, $\bar E_3$, $\bar E_8$,
and $\cos \theta$. One can express the imaginary part in terms of
three Casimir invariants in a similar manner as in the previous
section
\bea
\cE_p^2 &=& - \sqrt{\dfrac{C_4}{3}} \sin \phi_p +
\dfrac{2}{3} C_2,
\eea
with values of $\phi_p$ given by the same
Eqn. (\ref{cubsol}). A special case when $\cos \theta = 1$ has been
considered in \cite{nayak}.

We can obtain a general expression for the total production rate
from (\ref{eq:quark3}) with the $p_T$-integral,
\bea
Im~\Delta {\cal
L}_{q}|_{tot}=\dfrac{g^2}{32 \pi^3} \sum_p \sum_{n=1}^\infty
\dfrac{{\cal E}_p^2}{n^2}
 \exp\big(-\dfrac{2\pi n m^2}{g{\cal E}_p} \big).
\label{eq:quarktot}
\eea
 We plot the imaginary part (\ref{eq:quarktot})
for two values of the angle parameter $\theta$ in Fig. 3 for
comparison.
\begin{figure}[t]
\begin{center}
\psfig{figure=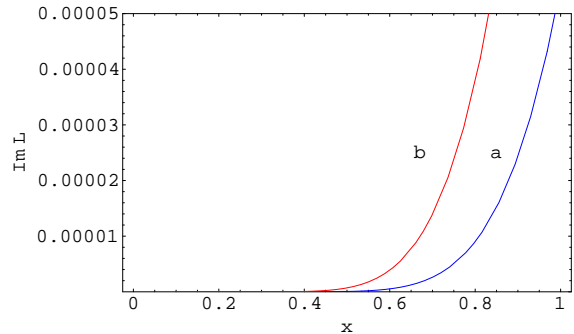, height = 4.5 cm, width = 7.5 cm}
\end{center}
\caption{\label{Fig. 3} The quark production rate:
(a) in quark-gluon plasma with $\cos \theta=0$
and (b) in hadron collider with $\cos \theta=1$.
Here we put $g \bar E_3=g \bar E_8=x$, and $m=1$.}
\end{figure}

The QCD effective action has been considered before
with different methods \cite{savv,yil,sch,ambj1,ditt,flyv,avr}.
Our method has the advantage that it naturally reduces
the calculation of $SU(N)$ QCD effective action
to that of $SU(2)$ QCD, and
we provide an explicit expression for
the effective action in terms of three gauge invariant Casimir
quantities for the most general constant background.
In most previous approaches only a special type of constant background
with one vector field component has been used.
Obviously, such a limitation can not provide
correct results in some physical applications.

We emphasize, however, that although the quark production
rate depends on three variables in general,
the actual number of independent
variables depends on case by case. For example, in hadron colliders
two chromoelectric fluxes $F_{\mu\nu}^3$ and $F_{\mu\nu}^8$ in
head-on collisions have the same direction,
the beam direction, so that we have to
put $\cos \theta=1$ \cite{cash,nayak}.
On the other hand, for the quark-gluon plasma
in the early universe or in astrophysics
we should average the angle $\theta$, because two chromoelectric
fluxes in such cases have no correlation in general.

{\bf Acknowledgements}

One of authors (DGP) thanks Prof. N.I. Kochelev for interesting
discussions. The work is supported in part by the ABRL Program of
Korea Science and Engineering Foundation (R14-2003-012-01002-0).

\end{document}